\documentclass[aoas,MSNbibl,nameyear,seceqn,dvips]{arximspdf}
\RequirePackage{algpseudocode}
\usepackage{algorithm}
\usepackage{graphicx}

\doi{10.1214/12-AOAS611} 
\volume{7}
\issue{2}
\pubyear{2013}
\firstpage{955}
\lastpage{988}

\makeatletter
\newcommand{\rrvert}{\vert}
\newcommand{\llvert}{\vert}
\makeatother

\begin{document}
\begin{frontmatter}

\title{Dynamic filtering of static dipoles in
magnetoencephalography}
\runtitle{Filtering of static dipoles}

\begin{aug}
\author{\fnms{Alberto}~\snm{Sorrentino}\corref{}\thanksref{t1}\ead[label=e1]{A.Sorrentino@warwick.ac.uk}},
\author{\fnms{Adam~M.}~\snm{Johansen}\thanksref{t2}\ead[label=e2]{A.M.Johansen@warwick.ac.uk}},
\author{\fnms{John~A.~D.}~\snm{Aston}\thanksref{t3}\ead[label=e3]{J.A.D.Aston@warwick.ac.uk}},
\author{\fnms{Thomas~E.}~\snm{Nichols}\thanksref{t4}\ead[label=e4]{T.E.Nichols@warwick.ac.uk}}\break
\and
\author{\fnms{Wilfrid~S.}~\snm{Kendall}\ead[label=e5]{W.S.Kendall@warwick.ac.uk}}
\thankstext{t1}{Supported in part by a Marie Curie Intra European Fellowship
within the 7th European Community Framework Programme.}
\thankstext{t2}{Supported in part by Engineering and Physical
Sciences Research Council Grant EP/I017984/1.}
\thankstext{t3}{Supported in part by Engineering and Physical
Sciences Research Council Grant EP/H016856/1 as well as the
EPSRC/HEFCE CRiSM grant.}
\thankstext{t4}{Supported in part by Medical Research Council Grant G0900908.}
\runauthor{A. Sorrentino et al.}
\affiliation{University of Warwick}
\address{Department of Statistics\\
University of Warwick\\
Coventry\\
CV4 7AL\\
United Kingdom\\
\printead{e1}\\
\phantom{E-mail:\ }\printead*{e2}\\
\phantom{E-mail:\ }\printead*{e3}\\
\phantom{E-mail:\ }\printead*{e4}\\
\phantom{E-mail:\ }\printead*{e5}} 
\end{aug}

\received{\smonth{5} \syear{2012}}
\revised{\smonth{9} \syear{2012}}


\begin{abstract}
We consider the problem of estimating neural activity from measurements
of the magnetic fields recorded by magnetoencephalography. We exploit the
temporal structure of the problem and model the
neural current as a collection of evolving current dipoles, which
appear and disappear, but whose locations are
constant throughout their lifetime. This fully reflects the
physiological interpretation of the model.

In order to conduct inference under this proposed model, it was
necessary to develop an algorithm based around state-of-the-art
sequential Monte Carlo methods employing carefully designed importance
distributions. Previous work employed a bootstrap filter and an
artificial dynamic
structure where dipoles performed a random walk in space, yielding
nonphysical artefacts in the reconstructions; such artefacts are not
observed when using the proposed model. The algorithm is validated with
simulated data, in which it provided an average
localisation error which is approximately half that of the bootstrap filter.
An application to complex real data derived from a somatosensory
experiment is presented. Assessment of model fit via marginal
likelihood showed a clear preference for the proposed model and the
associated reconstructions show better localisation.
\end{abstract}

%

\end{frontmatter}\vfill

\section{Introduction}\label{sec1}
Magnetoencephalography (MEG) [\citeauthor{haetal93}\break (\citeyear{haetal93})] is an imaging technique which
uses a helmet-shaped array of superconducting sensors to measure,
noninvasively, magnetic fields produced by underlying neural currents
in a human brain.
The sampling rate of MEG recordings is typically around 1000 Hz, which
allows observation of
neural dynamics at the
millisecond scale. Among
other noninvasive neuroimaging tools, only
electroencephalography (EEG) features a comparable temporal resolution.
EEG can
be considered complementary to MEG, due to its different sensitivity to source
orientation and depth [\citet{cocu83}].
Note that
estimation of the neural
currents from the measured electric or magnetic fields is an ill-posed inverse
problem [\citet{sa87}]: specifically, there are infinitely many possible
solutions,
because there exist source configurations that do not produce any detectable
field outside the head.

There are two well-established approaches to source modeling of MEG
data. In the \textit{distributed source} approach, the neural current is
modeled as a continuous vector field inside the head, discretized on a
large set of voxels; in this
case, the inverse problem is linear, and standard regularization
algorithms can be applied: commonly used methods include Minimum Norm Estimation
[\citeauthor{hail84} (\citeyear{hail84,hail94})], a Tikhonov-regularized solution corresponding
to the
Bayesian maximum a posteriori (MAP) solution under a Gaussian prior,
Minimum Current
Estimation (MCE) [\citet{uuhaso99}], an $L^1$
minimization that corresponds to the MAP associated with an exponential prior
in the Bayesian framework,
and beamforming [\citet{vvetal97}]. In this work we use the other
approach, a \textit{current dipole} model, where neural current is
modeled as a small set of point sources, or current dipoles;
each dipole represents the activity of a small patch of the brain cortex
as an electric current concentrated at a single point. A current dipole
is a six-dimensional
object: three coordinates define the dipole location within the brain,
a further
three coordinates define the dipole orientation and strength (the
\textit{dipole moment}).
The dipole model is a useful low-dimensional representation of brain activity:
in typical MEG experiments aimed at studying the brain response to external
stimuli [\citet{maetal97,sccr86}], the neural activity is modeled with
a very small number of dipoles, whose locations
are fixed but which have activity that evolves over time. However,
estimation of dipole parameters is mathematically more challenging than
estimation of the whole vector field, for at least two reasons: first,
the number
of dipoles is generally unknown and must be estimated from the data;
second, the
measured signals depend nonlinearly on the dipole location. For these two
reasons, dipole estimation is still largely performed with simple nonlinear
optimization algorithms that have to be initialized and supervised by
expert users,
although a few more advanced methods exist [\citet{mole99,juetal05}].

Distributed source methods are more prevalent, and most of them
estimate the
source distribution independently at each time point. However, since
the time interval
between two subsequent recordings is so small---about one millisecond---the
underlying brain activity does not much change between consecutive
measurements. Spatio-temporal modeling can incorporate this prior
knowledge by requiring that the solution
satisfy some form of temporal continuity. The availability of
increasing computational resources has made it possible to explicitly account
for the dynamic nature of the problem;
\citet{ouhago09}, \citet{tili11}, \citet{grkoha12} and \citet{tietal12}
employ spatio-temporal regularisation.

Recently, MEG source estimation has been cast as a filtering problem
within a
state-space modeling framework.
This approach has the further appeal that, in principle, it can be
used on-line, producing sequential updating of the posterior
distribution that incorporates the new data as they become available
at a computational cost (per measurement update) which does not increase
unboundedly over time.
In Long et~al. (\citeyear{loetal06,loetal11}), a distributed source model was
used and inference obtained with a high-dimensional Kalman filter. In
\citet{sovoka03}, \citet{caetal08}, \citet{soetal09} and \citet{paetal10} a
dipole model was used, and
the posterior distribution of the multi-dipole configuration was approximated
either with a bootstrap or with an approximation to
a Rao--Blackwellized bootstrap particle filter; however, the nature of
the approximation to the
Rao--Blackwellized filter was such that it yields underestimated uncertainty.
However, in the interests of computational expediency, all of these
studies employed an
artificial dynamic model in which dipole locations were modeled
as performing a random walk in the brain.

In this study we present a novel state-space model for MEG data,
based on current dipoles.
Unlike most other work in this area, the proposed approach explicitly
models the number of dipoles as a random variable, allowing new dipoles to
become active and existing dipoles to stop producing a signal.
In contrast to previous work on Bayesian filtering of multi-dipole models,
we treat the location of a dipole source as fixed over the lifetime of the
source. This is
in accordance with the general neurophysiological interpretation of a dipole
as
arising from the coherent activity of neurons in a small patch of cortex.
This is not a minor modification: it significantly influences the results,
their interpretation and the computational techniques which are
required in
order to perform inference.
The fact that dipole locations do not change over time
raises a computational challenge: while it would seem natural to adopt a
sequential Monte Carlo algorithm to approximate the posterior
distribution for
our state-space model, it is well known that these methods are not well suited
to the direct estimation of static parameters
although a number of algorithms have been developed to address this particular
problem in recent years [\citet{kaetal09}].
Standard particle filters are well suited to the estimation of time-varying
parameters in ergodic state-space models, as they can exploit knowledge
of the dynamics of the system itself to provide a good exploration of the
parameter space.
In order to
perform inference for the proposed model effectively, we adopt a
strategy based upon
the Resample-Move algorithm [\citet{gibe01}]: we introduce a Markov Chain
Monte Carlo move, formally targeting the whole posterior distribution
in the path space,
as a means to explore the state space while working with near-static
parameters. We note that the appearance and disappearance of dipoles provides
some level of ergodicity and ensures that there are no truly static parameters
within the state vector; this also implies that algorithms
appropriate for the estimation of true static parameters are not
applicable in
the current context. The proposed dynamic structure is exploited to
allow us to
implement MCMC moves on this space which mix adequately for the estimation
task at hand without having computational cost which increases unboundedly
as more observations become available.
In addition, we improve the basic importance sampling step with the introduction
of a carefully designed importance distribution.

The remainder of this paper has the following structure: Section
\ref{secfiltering} provides a very brief summary of filtering in
general and
particle filtering in particular, Section \ref{secmethodology}
introduces the proposed
models and associated algorithms, and Sections \ref{secsimulation} and
\ref{secexample}
provide validation of these algorithms via a simulation study and an
illustration of performance on real data. A brief discussion is
provided in
the final section.

\section{Bayesian and particle filtering}
\label{secfiltering}
Bayesian filtering is a general approach to Bayesian inference for
Hidden Markov models: one is interested in the sequence of posterior
distributions $\{p(j_{0:t}|b_{1:t})\}_{t=1,\ldots,T}$, and particularly the
associated marginal distributions $\{p(j_t|b_{1:t})\}_{t=1,\ldots,T}$,
for the unobserved process $\{J_1,\ldots, J_t,\ldots\}$ given
realizations of the
measurements $\{B_1,\ldots, B_t,\ldots\}$, where $j_t$ and $b_t$ are
instances of the corresponding random variables. If one assumes that:

\begin{itemize}
\item the stochastic process $\{J_t\}$ is a first order Markov process with
initial distribution $p(j_0)$ and homogeneous transition probabilities
$p_t(j_{t+1}|j_t) = p(j_{t+1}|j_t)$, such that $p(j_{0:t}) = p(j_0)
\prod_k p(j_{k+1}|j_k)$; in MEG, this corresponds to the model for
evolution of current dipoles;
\item each observation $B_t$ is statistically
independent of the past observations given the current state $j_t$, and has
conditional distribution $p_t(b_t|j_t)$, which it is convenient to
treat as
time homogeneous, $p_t(b_t|j_t) = p(b_t|j_t)$; in MEG, the observations
are thus assumed to only depend on the current neural configuration.
\end{itemize}
Then the posterior distribution at time $t$ is given by
%
\begin{equation}
p(j_{0:t}|b_{1:t}) = \frac{p(j_{0:t},b_{1:t})}{p(b_{1:t})} = \frac{p(j_0) \prod_{k=1}^t p(j_{k}|j_{k-1})
p(b_k|j_k)}{p(b_{1:t})},
\label{fullbayes}
\end{equation}
and satisfies the recursion
%
\begin{equation}
\label{recursion} p(j_{0:t}|b_{1:t}) = p(j_{0:t-1}|b_{1:t-1})
\frac{p(b_t|j_t)p(j_t|j_{t-1})}{p(b_t|b_{1:t-1})}.
\end{equation}

Unfortunately, this recursion is only a formal solution, as it is not possible
to evaluate the denominator except in a few special cases, notably linear
Gaussian and finite state-space models. It is, therefore, necessary to resort
to numerical approximations to perform inference in these systems.
Particle filters [see \citet{gosasm93}, \citet{caclfe99}, and \citet
{gibe01} for original work of particular
relevance to the present paper and \citet{dojo11} for a recent review]
are a class of methods
that combine
importance sampling and
resampling steps in a sequential framework, in order to obtain samples
approximately distributed according to
each of the filtering densities in turn.
These algorithms are especially well suited to applications in which a
time-series of measurements is available and interest is focussed on
obtaining
on-line
updates to
the
information about the
current state of the
unobservable system---such as the current neural activity in the
context of MEG.

Importance sampling is one basic element of particle filtering: it is a
standard technique for approximating the expectation $\int f(x) p(x)
\,dx$ of a reasonably well-behaved function $f$
under a density $p(x)$ when i.i.d. samples from $p(x)$ are unavailable;
the strategy consists in sampling $\{x^i\}_{i=1,\ldots,N}$ from an
\textit{importance} density $q(x)$ such that $q(x)/p(x) > 0$, and then
approximating
%
\begin{equation}
\int f(x)p(x)\,dx = \int f(x) \frac{p(x)}{q(x)} q(x) \,dx \simeq \sum
_i f\bigl(x^i\bigr) w^i,
\end{equation}
where the weights $w^i \propto p(x^i)/q(x^i)$ correct for the use of
the importance density
and are normalized such that $\sum_i w^i = 1$.
Developing good proposal distributions for the MEG setting is one contribution
of the present paper.
Conditions of boundedness of the weight function $p(x)/q(x)$, and
finiteness of
the variance of $f(X)$ for $X \sim p(\cdot)$, are together sufficient
to ensure that this estimator obeys a central limit
theorem with finite variance [\citet{ge89}].

To apply importance sampling to the posterior density, one could
sample $N_{\mathrm{particles}}$
points (or \textit{particles}) $\{j_{0:t}^1,\ldots,j_{0:t}^{N_{\mathrm{particles}}}\}$ from a
proposal density $q(j_{0:t}|b_{1:t})$ and associate a weight $w_t^i
\propto
\frac{p(j_{0:t}^i|b_{1:t})}{q(j^i_{0:t}|b_{1:t})}$ to each particle.
In the sequential framework, importance sampling can
be simplified by a proper choice of the importance density: if the
importance density is
such that $q(j_{0:t}|b_{1:t}) = q(j_0) \prod_k q(j_k|j_{1:k-1},b_k)$,
then given
the sample set at time $t-1$,
$\{j_{0:t-1}^1,\ldots, j_{0:t-1}^{N_{\mathrm{particles}}}\}$,
which is appropriately weighted to approximate $p(j_{0:t-1}|b_{1:t-1})$,
one can
draw $\{j_t^i\}$ from $q(j_t|j_{0:t-1}^i, b_t)$ and set $j_{0:t}^i =
(j_{0:t-1}^i,
j_t^i)$. Furthermore, thanks to the recursion (\ref{recursion}), one
can update the particle weight recursively,
%
\begin{eqnarray}
w_t^i &\propto & \frac{p(j^i_{0:t}|b_{1:t})}{q(j^i_{0:t}|b_{1:t})} = \frac{p(j_{0:t-1}^i|b_{1:t-1}) p(j_t^i|j_{t-1}^i) p(b_t|j_t^i) /
p(b_{t}|b_{1:t-1})}{q(j_{0:t-1}^i) q(j_t^i|j_{t-1}^i,b_t)}
\nonumber
\\[-8pt]
\\[-8pt]
\nonumber
&\propto & w_{t-1}^i \frac{p(b_t|j^i_t) p(j^i_t|j^i_{t-1})}{ q(j^i_t|j^i_{t-1}, b_t)}.
\end{eqnarray}

Resampling is a stochastic procedure
which attempts to address the inevitable increase in the
variance of the importance sampling estimator as the length of the time series
being analysed increases
by replicating particles with large weights and eliminating those with
small weights.
The expected number of ``offspring'' of each particle is
precisely the product of $N_{\mathrm{particles}}$ and its weight before
resampling. The
unweighted population produced by resampling is then propagated forward
by the
recursive mechanism described above.
See \citet{doca05} for a comparison between some of the most
common approaches. In the experiments below the \textit{systematic} resampling
scheme of \citet{caclfe99} was employed.
The resampling step allows good approximations of the filtering density
$p(j_t|b_{1:t})$ to be maintained and helps to control the variance of the
estimates over time [\citet{ch04,delm04}].

One issue in the application of particle filtering is the choice of
the importance distribution. The simplest choice, leading to the so-called
\textit{bootstrap} filter, is to use the prior distribution as an importance
distribution, setting $q(j_t|j_{0:t-1},b_t) = p(j_t|j_{t-1})$.
However, when the likelihood is informative, this choice will lead to an
extremely high variance estimator. A good importance
density should produce reasonably uniform importance weights,
that is, the variance of the importance weights should be small.
The optimal importance distribution that minimizes the
conditional variance of the importance weights, whilst factorising
appropriately, is given [\citet{dogoan00}]
by
%
\begin{equation}
q^\star(j_t|b_t,j_{t-1}) =
p(j_t|b_t,j_{t-1}) = \frac{p(j_t|j_{t-1})p(b_t|j_t)}{p(b_t|j_{t-1})};
\label{optimalimportance}
\end{equation}
in practice, one should always try to approximate this distribution as
well as
is computationally feasible.
Furthermore, a convenient way to monitor the variance of the importance weights
is by looking at the effective sample size [\citet{koliwo94}, ESS],
defined as
%
\begin{equation}
\mathrm{ESS} = \Biggl(\sum_{i=1}^N
\bigl(w_t^i\bigr)^2 \Biggr)^{-1}
\label{ess}.
\end{equation}
The ESS ranges between 1 and $N_{\mathrm{particles}}$
and can be thought of as an estimate (good only when the
particle set is able to provide a good approximation of the posterior density
of interest) of the number of independent samples from the posterior which
would be expected to produce an estimator with the same variance as the
importance sampling estimator which was actually used (resampling somewhat
complicates the picture in the SMC setting).

It is possible to obtain an estimate
of the marginal likelihood (which, remarkably, is unbiased [\citet{delm04}])
from the particle filter output,
%
\begin{equation}
\label{eqmarginallikelihood} p(b_{1:t}) = p(b_1) \prod
_{n=1}^{t-1} p(b_{n+1}|b_{1:n}),
\end{equation}
using the direct approximation of the \textit{conditional likelihood},
%
\begin{equation}
p(b_{n+1}|b_{1:n}) \approx\sum_i
\tilde{w}^i_{n+1} \Rightarrow p(b_{1:t}) \approx
\prod_{n=1}^t \sum
_i \tilde{w}_{n}^i,
\end{equation}
where $\tilde{w}^i_{n+1}$ are the unnormalized weights at time $n+1$.\

\section{Filtering of static dipoles}
\label{secmethodology}
\subsection{Statistical model}

In MEG, we are given a sequence of recordings of the magnetic
field $\{b_t\}_{t=1,\ldots,T}$, and wish to perform inference on the underlying
neural current $\{j_t\}_{t=1,\ldots,T}$ that has produced the measured fields.

In this study we model the neural current as a set of current dipoles
$j_t = \{d_t^{(1)},\ldots,d_t^{(N_t)}\}$; here and
in the rest of the paper, superscripted parenthetical indices label
individual dipoles within a dipole set.
Each current dipole $d_t^{(i)} = (r_t^{(i)}, q_t^{(i)})$ is
characterized by a
location $r_t^{(i)}$, within the brain volume, and a dipole moment $q_t^{(i)}$,
representing direction and strength of the neural current at that location.

In order to perform Bayesian inference, we need to specify two
distributions: the prior distribution for the neural current in time
and the likelihood function.

\subsubsection*{Prior distributions}
We specify the prior distribution for the spatio-temporal evolution of the
neural current by providing the prior distribution at $t=0$ and a
homogeneous transition kernel $p(j_t|j_{t-1})$. We devise our prior
model for
the neural current path following two basic principles:
\begin{itemize}
\item at any time point $t$, the number of active dipoles $N_t$ is
expected to
be small and the average dipole lifetime is around 30
milliseconds;\eject
\item dipole moments change (continuously) in time, to model
increasing/dimin\-ishing activity of
a given neural population, but dipole locations are fixed; for this reason,
we term this the Static model.
\end{itemize}

In addition, for computational reasons we impose an upper bound on the number
of simultaneously active dipoles $N_t$: in the experiments below we set
this upper bound to $N_{\mathrm{max}}=7$,
as our informal prior expectation on the number of dipoles is markedly
less than 7,
and this is born out by the data.
Finally, for both computational and modeling reasons, dipole locations
are required to belong to a finite set of predefined values $r^{(i)}
\in R_{\mathrm{grid}}$ with $R_{\mathrm{grid}} = \{r^k_{\mathrm{grid}}\}_{k=1}^{N_{\mathrm{grid}}}$.
It is customary in MEG research to use this kind of grid,
where points are distributed along the
cortical surface, the part
of the brain where the neural currents flow.
At the computational level, the use of these grids allows
precalculation of the forward problem, that is, of the
magnetic field produced by unit dipoles, as described later.
Automated routines for segmentation and reconstruction of the cortical
surface from Magnetic Resonance images have been available
for over ten years. In the experiments below we used Freesurfer [\citet
{dafise99}] to obtain the tessellation of the cortical
surface from the Magnetic Resonance data. We then used the MNE software
package (\url{http://www.martinos.org/mne/}) to
get a subsample of this tessellation with a spacing of 5 mm; the
resulting grid contains 12,324 distinct locations.

At time $t=0$ the initial number of dipoles $N_0$ is assumed to follow
a truncated
Poisson distribution with rate parameter $1$ and maximum $N_{\mathrm{max}}$;
we then specify a uniform distribution over the grid points for the
dipole locations, and a Gaussian distribution for the dipole moments,
leading to the joint prior distribution:
%
\begin{equation}
\label{eqprior} p(j_0) = \sum_k
P(N_0 = k) \prod_{n=1}^k
U_{R_{\mathrm{grid}}}\bigl(r_0^{(n)}\bigr) \mathcal{N}
\bigl(q_0^{(n)}; 0, \sigma_q \mathbf{I}\bigr),
\end{equation}
where $U_{R_{\mathrm{grid}}}(\cdot)$ is the uniform distribution over the
set $R_{\mathrm{grid}}$
and $\mathcal{N}(\cdot;\mu, \Sigma)$ is the Gaussian density of mean
$\mu$ and covariance $\Sigma$.

The transition density accounts for the possibility of dipole birth and dipole
death, as well as for the evolution of individual dipoles. It is
assumed that
only one birth or one death can happen at any time point. The
transition density is composed of three summands as follows:
%
\begin{eqnarray}
\label{eqtransition} && p(j_t | j_{t-1})\nonumber
\\
&&\qquad =P_{\mathrm{birth}} \times U_{R_{\mathrm{grid}}}\bigl(r_t^{(N_t)}
\bigr) \mathcal{N}\bigl(q_t^{(N_t)};0, \Delta\bigr) \times
\prod_{n=1}^{N_{t-1}} \delta_{r_t^{(n)},r_{t-1}^{(n)}}
\mathcal{N} \bigl(q_t^{(n)}; q^{(n)}_{t-1},
\Delta\bigr)
\nonumber
\\[-8pt]
\\[-8pt]
\nonumber
&&\qquad\quad{}+ P_{\mathrm{death}} \times \frac{1}{N_{t-1}} \sum_{j=1}^{N_{t-1}}
\prod_{n=1}^{N_{t-1}-1} \delta_{r_t^{(n)},
r_{t-1}^{(a_{j,n})}}
\mathcal{N} \bigl(q_{t}^{(n)}; q_{t-1}^{(a_{j,n})},
\Delta\bigr)
\\
&&\qquad\quad{}+ (1-P_{\mathrm{birth}} - P_{\mathrm{death}}) \times \prod
_{n=1}^{N_{t-1}} \delta_{r_t^{(n)},r_{t-1}^{(n)}} \mathcal{N}
\bigl(q_t^{(n)};q^{(n)}_{t-1},\Delta\bigr),\nonumber
\end{eqnarray}
where $\delta_{\cdot,\cdot}$ is the Kronecker delta function. The first
term in equation (\ref{eqtransition}) accounts for the
possibility that a new dipole appears, with probability $P_{\mathrm{birth}}$; the
location of the new dipole, for convenience the $N_t$th dipole of the set,
is uniformly distributed in the brain, while the dipole moment has a Gaussian
distribution. All other dipoles evolve independently: dipole locations remain
the same as in the previous time point, while dipole moments perform a Gaussian
random walk.
The second summand in equation (\ref{eqtransition}) accounts for the
possibility that one of the existing dipoles disappears: one of the
dipoles in
the set at time $t-1$ is excluded from the set at time $t$; all existing
dipoles have equal probability of disappearing; surviving dipoles evolve
according to the same rules described earlier. The disappearance of a dipole
entails a rearrangement of the dipole labels, namely, the label of a dipole
changes if a dipole with a smaller label disappears from the set. Here
$a_{j,n}$ is the label
of the ancestor of the $n$th dipole after the death of the $j$th
dipole, and is given by
%
\begin{equation}
a_{j,n} = \cases{ n, &\quad $\mbox{if } n<j,$
\cr
n+1, &\quad  $\mbox{if } n \ge j$.}
\end{equation}
Finally, in the last term the number of dipoles in the set remains the same.

The parameters of these prior distributions were set to reflect our informal
(and
neurophysiologically-motivated)
prior expectations for the number of dipoles and their temporal evolution.
Birth and death probabilities were set, respectively, to $P_{\mathrm{birth}}
= 1/100$ and $P_{\mathrm{death}} = (1- (1-1/30)^{N_t})$,
as the expected lifetime of a single dipole is about $30$ time points,
since simultaneous deaths are
neglected. In addition, due to the presence of an upper bound to the
number of simultaneous dipoles, the birth probability
is zero when $N_t$ is equal to $N_{\mathrm{max}}$. Inference is
insensitive to
the precise value of $N_{\mathrm{max}}$ provided that it is
sufficiently large.
In simulation experiments we found that estimation was robust to moderate
changes in these parameter values, as long as they remained compatible
with the assumption
that the number of sources is small. As a consequence of depending upon
a finite sample approximation of the posterior,
better estimation of the precise time of dipole disappearance could be
obtained by increasing the death
probability to a substantially larger value. However,
such large death probabilities would render the prior average dipole
lifetime unrealistically short and,
thus, we preferred to use a value that makes our prior\vadjust{\goodbreak} as close as
possible to
the underlying physiological process.
The transition density for the dipole moment is Gaussian, but the covariance
matrix is not isotropic:
the variance is ten times larger in the direction of the
dipole moment itself, thus giving preference to changes in strength
relative to changes in the orientation.

\subsubsection*{Likelihood}
The magnetic field distribution is measured by an array of SQUID-based
(Superconducting
QUantum Interference Device) sensors, arranged around the subject's
head producing,
at time $t$, a column vector $b_t$ containing one entry for each of the
$N_{\mathrm{sensors}}$ sensors.

The relationship between the neural current parameters and the
experimental data
is contained in the \textit{leadfield} or \textit{gain} matrix $G$.
The size of the leadfield matrix is $N_{\mathrm{sensors}} \times3 N_{\mathrm{grid}}$: each column
contains the magnetic field produced by a unit dipole placed in one of
the $N_{\mathrm{grid}}$ grid
points and oriented along one of the three orthogonal directions.
Calculation of the leadfield matrix involves the simulation of how the
electromagnetic fields propagate inside the subject's head, hence
requiring as accurate as possible models of the conductivity profile
inside the head. In the experiments
below, we used a standard 4-compartment model, comprising the brain,
the cerebro-spinal fluid, the skull and
the scalp; the boundaries of these compartments were extracted from the
Magnetic Resonance images of the subject,
and the Boundary Element method implemented in MNE was used to
calculate the leadfield.
We denote by $G(r^k)$ the matrix of size $N_{\mathrm{sensors}} \times3$
that contains the fields produced by
the three orthogonal dipoles at $r^k$. The measurement model is
%
\begin{equation}
\label{biotsavart} b_t = \sum_{i=1}^{N_t}
G\bigl(r_t^{(i)}\bigr) q_t^{(i)} +
\varepsilon_t,
\end{equation}
where $\varepsilon_t$ is an additive noise vector.
Assuming that the $\varepsilon_t$
are independent and Gaussian with covariance $\Sigma_{\mathrm{noise}}$ leads
to the likelihood,
%
\begin{equation}
\label{eqlikelihood} p(b_t|j_t) = \mathcal{N} \Biggl(
b_t; \sum_{i=1}^{N_t} G
\bigl(r_t^{(i)}\bigr) q_t^{(i)},
\Sigma_{\mathrm{noise}} \Biggr).
\end{equation}

\subsection{Computational algorithm}

The principal difficulty with the development of effective particle filtering
algorithms for the static model described in the previous section is
as follows: the dipole locations, except at the times of appearance and
disappearance, behave
as static parameters. Standard sequential Monte Carlo
algorithms operating as filters/smoothers are not well suited to
inference in
the presence of static parameters and a variety of techniques have been
devised to address that particular problem [\citet{kaetal09}].
If one has a Hidden Markov model with unknown static parameters,
if one simply augments the state vector with the unknown static
parameters and introduces an\vadjust{\goodbreak}
additional degenerate element in the transition kernel, then one quickly
suffers from degeneracy---in the case of the sequential importance
resampling algorithm, for example, the algorithm is dependent upon
sampling good values for the static parameters at
the beginning of the sampling procedure, as there is no mechanism for
subsequently introducing any additional diversity.
The problem is exacerbated by the fact that this
initial sampling stage is extremely difficult in the MEG context, as
the state
space is large and the relationship between the likelihood and the underlying
states is complex and nonlinear. Below, we develop strategies which exploit
the fact that the dipole locations are not really static parameters, as they
persist for only a random subset of the time sequence being analysed, together
with more sophisticated sequential Monte Carlo techniques.

Here we propose a computational algorithm characterized by two main features:
First, a mechanism
that exploits the Resample-Move idea
[\citet{gibe01}]
in order to mitigate considerably
the degeneracy effect
produced by the static parameters; the idea is to introduce a Markov
Chain Monte Carlo move at
each iteration, targeting the whole posterior distribution.
Second, a well designed importance distribution, in which birth
locations and
deaths are drawn from approximations to the optimal importance density
(\ref{optimalimportance}).

\subsubsection*{Resample-Move}

The Resample-Move algorithm is an approach for addressing degeneracy
in sequential Monte Carlo algorithms. The idea is to use a Markov kernel
$K(j_{0:t}^{\prime}|j_{0:t})$ of invariant distribution $p(j_{0:t}|b_{1:t})$
to provide diversity in the sample set. The underlying computational machinery
is still sequential importance resampling and its validity does not depend
upon the ergodicity of Markov chains. If $J_{0:t}$ is distributed
according to $p(j_{0:t}|b_{1:t})$, and $J_{0:t}'|J_{0:t}$ is distributed
according to $K(j_{0:t}^{\prime}|j_{0:t})$, then $J_{0:t}'$ is still marginally
distributed according to $p(j_{0:t}|b_{1:t})$ and, more generally, the
distribution of $J_{0:t}^\prime$ cannot
differ more from $p(j_{0:t}|b_{1:t})$ in total
variation than does the distribution of $J_{0:t}$.

In this study, the Markov kernel is constructed following the standard
Metropolis Hastings algorithm: proposed samples $j_{0:t}^{\prime}$ are drawn
from a
proposal distribution $L(j_{0:t}^{\prime}\phantom{}|j_{0:t})$ and then accepted with
probability $\alpha$, with
%
\begin{equation}
\label{acceptance} \alpha= \min \biggl( 1, \frac{p(j'_{0:t}|b_{1:t})
L(j_{0:t}|j_{0:t}')}{p(j_{0:t}|b_{1:t}) L(j_{0:t}'|j_{0:t})} \biggr).
\end{equation}
Since the purpose of this move is to introduce diversity for the dipole
locations, we devised a simple proposal distribution that involves only a
modification of the dipole locations, modifying one dipole at a time.
Specifically,
at every time point and for each particle we propose sequentially $N_t^i$
moves, where $N_t^i$ is the number of dipoles in the particle: for each dipole
we choose one of its neighbours at random (one of the grid points
within a fixed
radius of 1~cm);
the proposed particle $j_{0:t}'$ differs from the original particle
$j_{0:t}$ only in the location of the proposed dipole; the acceptance
probability $\alpha$
is dominated by the
ratio of the likelihood of the original and the displaced particles,
that can
only differ for time points after the appearance of the dipole at (say)
time $t=t_0$,
%
\begin{equation}
\alpha= \min \biggl( 1, \frac{|S|}{|S^\prime|}\frac{\prod_{n=t_0}^{t}
p(b_n|j_n')}{\prod_{n=t_0}^{t} p(b_n|j_n)} \biggr),
\label{rmacceptance}
\end{equation}
where $|S|$ is the number of neighbours of the dipole in $j_{0:t}$ and
$|S^\prime|$
is the number of neighbours of the dipole in $j_{0:t}'$. Note that the
$|S|/|S^\prime|$ factor arises from the asymmetric proposal---although it
may, initially, appear symmetric, the restriction to an irregular
discretisation grid induces asymmetry.
\subsubsection*{Importance distribution}

As mentioned in Section \ref{secfiltering}, having a good importance
distribution is important in order to make a particle filter work in
practice. At the same time, in our case
the optimal importance distribution (\ref{optimalimportance})
is intractable---as it generally is for realistic models.
Here we propose an importance density that features an acceptable
computational cost
but substantially improves the statistical efficiency at two crucial points.
When birth is proposed, instead of drawing uniformly from the brain,
the new
dipole location is sampled according to a heuristic distribution based
on the
data. Although the closeness of this distribution to the optimal importance
distribution will influence the variance of the estimator, the importance
sampling correction ensures that we obtain consistent (in the number of
particles) estimation under very mild conditions.
Conditional on not proposing a birth, a death is proposed
with approximately optimal probability. More precisely, we propose to
use the
following importance distribution:
%
\begin{eqnarray}
\label{importance} && q(j_t | j_{t-1}, b_t)\nonumber\\
&&\qquad =
Q_{\mathrm{birth}} \times q\bigl(r_t^{(N_t)},
q_t^{(N_t)}|b_t, j_{t-1}\bigr) \times
\prod_{n=1}^{N_{t-1}} \delta_{r_t^{(n)},r_{t-1}^{(n)}}
\mathcal{N} \bigl(q_t^{(n)}; q^{(n)}_{t-1},
\Delta\bigr)
\\
&&\qquad\quad{}+Q_{\mathrm{death}}(j_{t-1},b_t) \nonumber\\
&&\qquad\qquad{} \times \sum
_{j=1}^{N_{t-1}} P_{\mathrm{dying}}\bigl({d}^{(j)}|j_{t-1},
b_t\bigr)\times\prod_{n=1}^{N_{t-1}-1}
\delta_{r_t^{(n)}, r_{t-1}^{(a_{j,n})}} \mathcal{N} \bigl(q_{t}^{(n)};
q_{t-1}^{(a_{j,n})}, \Delta\bigr)
\nonumber\\
&&\qquad\quad{}+\bigl(1-Q_{\mathrm{birth}} - Q_{\mathrm{death}}(j_{t-1},b_t)
\bigr) \times\prod_{n=1}^{N_{t-1}}
\delta_{r_t^{(n)},r_{t-1}^{(n)}} \mathcal{N} \bigl(q_t^{(n)};q^{(n)}_{t-1},
\Delta\bigr).\nonumber
\end{eqnarray}

Birth is proposed at a fixed rate $Q_{\mathrm{birth}}$, because evaluating
the optimal birth probability
would require the evaluation of intractable integrals and even
obtaining a
reasonable approximation would be computationally prohibitive; we use
$Q_{\mathrm{birth}} = 1/3$ in our algorithm. In the absence
of a (near) optimal proposal, detecting new dipoles is the most challenging
task faced by the algorithm; it is therefore appropriate to dedicate
a substantial proportion of the computing resources to this task and so
we use a value
rather larger than $P_{\mathrm{birth}}=1/100$.
When birth is proposed, the new dipole
location is proposed from a heuristic proposal distribution $q(r_t^{(N_t)},
q_t^{(N_t)}|b_t, j_{t-1})$ computed from the data and obtained as
follows: consider the linear inverse
problem
%
\begin{equation}
b_t = G J_t + \varepsilon_t,
\end{equation}
where $G$ is the whole leadfield matrix and $J_t$ is a vector whose entries
$J_t^k = J_t(r_k)$ represent the current strength at each point $r_k$
of the
grid; this inverse problem can be solved with a Tikhonov
regularization method,
%
\begin{equation}
\widehat{J}_t = R G^T \bigl(G R G^T +
\lambda_{\mathrm{reg}} I\bigr) b_t,
\end{equation}
where $R$ is a weighting matrix which mitigates the bias toward
superficial sources and $\lambda_{\mathrm{reg}}$ is the regularization
parameter. In the experiments below, $R$ and $\lambda$ were chosen
according to the
guidelines given in \citet{lietal06}. The Tikhonov solution provides a
widespread estimate of neural
activity within the brain; by normalizing the Tikhonov solution, we
obtain a
spatial distribution which should
be largest in the regions in which there is the highest probability
that a
source is present:
%
\begin{equation}
\label{Tikhonovdensity} q(r|b_t) = \widehat{J}_t(r)/{
\sum_{r'} \widehat{J}_t
\bigl(r'\bigr).}
\end{equation}
Notice that the rescaled Tikhonov inverse used here is simply an importance
sampling proposal and the discrepancy between it and the posterior
distribution implied by the
Bayesian model is corrected for by importance weighting (and
resampling, as
required). Other heuristic inversion methods could be employed to provide
alternative proposal distributions.

This density does not depend on the actual particle state, which is a
significant computational advantage: it can be calculated just once per
iteration rather than once per particle per iteration. However, there
is a
drawback in that its performance is expected to worsen as the number
of dipoles
increases (and much of the mass associated with $q$ is located
close to existing dipoles).
We approximate the optimal death probability via an approximation in which
the dipole parameters do not change from $t-1$ to $t$:
death is proposed with probability
%
\begin{eqnarray}
\qquad&& Q_{\mathrm{death}} (j_{t-1},b_t)
\nonumber
\\[-8pt]
\\[-8pt]
\nonumber
&&\qquad =
\frac{(1-Q_{\mathrm{birth}}) \times1/N_{t-1} \sum_{k=1}^{N_{t-1}}
p(b_t|j_{t-1}^{(-k)}) P_{\mathrm{death}} }{1/N_{t-1}\sum_{k=1}^{N_{t-1}}
p(b_t|j_{t-1}^{(-k)}) P_{\mathrm{death}} + p(b_t|j_{t-1}) (1-P_{\mathrm{birth}} -
P_{\mathrm{death}})},
\end{eqnarray}
where $j_{t-1}^{(-k)} = \{d_{t-1}^{(1)},\ldots,d_{t-1}^{(k-1)},
d_{t-1}^{(k+1)},\ldots,d_{t-1}^{(N_{t-1})} \} $ is the dipole set
at time $t-1$ without the $k$th dipole;
if death is proposed, the dipole to be killed is drawn according to
%
\begin{equation}
P_{\mathrm{dying}}\bigl(d^{(k)}|j_{t-1}, b_t\bigr)
\propto p\bigl(b_t|j_{t-1}^{(-k)}\bigr).
\end{equation}
Otherwise, with probability $1 - Q_{\mathrm{birth}} - Q_{\mathrm{death}}(j_{t-1},b_t)$,
the number of dipoles remains the same.
The overall approach is outlined in Algorithm \ref{algrm}.

\begin{algorithm}[t]
\caption{Outline of the Resample-Move algorithm}\label{algrm}
\begin{algorithmic}
\For{$i=1,\ldots,N$}
\State draw $j_0^i$ from $p(j_0)$;
\EndFor

\For{$t=1,\ldots,T$}

\For{$i=1,\ldots,N$} (importance sampling)
\State draw $j_t^i$ from $q(j_t|\tilde{j}_{0:t-1}^i, b_t)$,
\State set $j_{0:t}^i = (j_{0:t-1}^i, j_t^i)$
\State compute the unnormalized weights $\tilde{w}_t^i = \frac
{p(b_t|j_t^i)p(j_{t}^i|j_{t-1}^i)}{q(j_t^i|j_{t-1}^i, b_t)}$
\EndFor

\For{$i=1,\ldots,N$} (normalize the weights)
\State$w_t^i = \tilde{w}_t^i / W_t$, with $W_t = \sum_i \tilde{w}_t^i$
\EndFor
\For{$i=1,\ldots,N$} (resample)
\State draw $\tilde{j}_{0:t}^i$ from $\{ j_{0:t}^i \}$, with
$P(\tilde{j}_t = j_t^k) = w_t^k\ \forall k$
\EndFor

\For{$i=1,\ldots,N$} (move)
\For{$k=1,\ldots,N_t^i$}
\State draw $r^\star$ from the neighbours of $r^{(k),i}_t$
\State accept the jump, replacing $r^{(k),i}_t$ with $r^\star$ with
probability given by
\State equation (\ref{rmacceptance})
\EndFor
\EndFor

\EndFor
\end{algorithmic}
\end{algorithm}

\subsection{Connections with previous work}

\label{subprevious}
Application of particle filtering for estimation of current dipole parameters
from MEG data has been described in \citet{caetal08}, \citet{caetal11},
\citet{paetal10}, \citet{soetal09} and \citet{so10}. A fundamental
difference between our work and previous studies is that they all used
a Random-Walk model, that is, dipole locations were allowed to change
in time, according to a random
walk. In addition, in previous studies birth and death
probabilities were set to $P_{\mathrm{birth}} = P_{\mathrm{death}} = 1/3$. The
computation was
performed with a standard bootstrap particle filter, but a heuristic
factor was used to penalize models with a large number of dipoles: the particle
weight, rather than being proportional to the likelihood alone, was in fact
proportional to $ \frac{1}{N^i_t!} p(b_t|j_t^i)$, where $N_t^i$ is the number
of dipoles.

Our proposed strategy has a number of benefits: it is fully Bayesian
and hence admits a clear
interpretation and, most importantly, the statistical model is
consistent with the biophysical
understanding of the system being modeled.
Experimentally, we found that models which incorporated
artificial dynamics (random-walk type models) led to significant
artefacts in
the reconstruction in which dipoles moved gradually
from one side of the brain to the
other in opposition to the interpretation of those dipoles as arising
from fixed neural
populations. Although the Resample-Move mechanism and Random-Walk
models may
appear superficially similar, they have very different interpretations and
consequences: using the Random-Walk model is equivalent to performing
inference under the assumption that the dipole location changes from one
iteration to the next; using the Resample-Move algorithm with the
Static model
leads to inference consistent with a model in which the dipoles do not move.

Below the Static model is compared with the Random-Walk model described in
previous studies; in our implementation of the Random-Walk model,
dipoles can
jump between neighbouring points, with a transition probability
proportional to $\exp(-d^2/2\sigma_d^2)$,
where $d$ is the distance between the two points and $\sigma_d = 0.5$
cm in the simulations below.

The use of improved importance distributions is also possible in the context
of the Random-Walk model and we have employed the importance distributions
described in the following section, which improved its performance in
comparison with the bootstrap approach employed previously.

\subsubsection*{Importance distributions for the Random-Walk model}

In the Random-Walk model, the transition probability
distribution
allows current dipole
locations to jump within the set of neighbouring points;
the use of bootstrap proposals to implement this, in conjunction with
random change of dipole moment,
will certainly result in a loss of sample points in the high
probability region, even in the course of a single step.
In our implementation of the Random-Walk model
we use the following approach in order to improve
importance sampling efficiency: for each dipole contained in the
particle---starting from
the one most recently born---we first sample the dipole moment
according to the dynamic model,
and then calculate the probability that a dipole with the sampled
dipole moment
is at any of the neighbouring locations, given the data and the other dipoles.
At each step the most recent values of the remaining parameters are
always used, hence, the $k$th dipole is
sampled conditional on the current values of the dipoles with a larger
label and on
the previous values of the dipoles with a smaller label.

The improved birth and death moves developed in the previous section
can also be
employed without modification in the Random-Walk model.

\subsection{Computational considerations}

We end this section with a very brief examination of various
computational aspects of the proposed algorithms.
In the MEG application,
likelihood evaluation is responsible for the vast majority of computational
effort. The only additional
computation in the proposed method apart from these evaluations is the
Tikhonov inverse solution, which is
quite fast, and is carried out once per iteration rather than once per
particle per iteration.
Because of this, the relative cost of the Tikhonov inverse computation
can be treated as negligible.
Consequently, we use the number of likelihood evaluations
required by the proposed algorithms
as a proxy for computational effort. We itemize this effort as follows:
\begin{itemize}
\item The total number of likelihood computations required by the
bootstrap filter
is $T N$, where $T$ is the total number of time points and
$N$ the number of particles.
\item The Resample-Move algorithm requires calculation of the
likelihood for the whole past history of each dipole, hence requiring an
additional\break $T N \bar{N}_{\mathrm{dip}} T_{\mathrm{life}}/2$,
where $T_{\mathrm{life}}$ is the average lifetime of a dipole.
\item The death proposal requires a number of additional likelihood
evaluations of
$TN\bar{N}_{\mathrm{dip}}$, where $\bar{N}_{\mathrm{dip}}$ is the average
number of dipoles.
\item Finally, for the Random-Walk model, the proposed conditional
importance sampling requires
the calculation of a number of likelihoods equal to the average number of
neighbours; this is done at every time step, for each active dipole, hence
bringing an additional cost of $TN\bar{N}_{\mathrm{dip}}N_{\mathrm{neighbours}}$.
\end{itemize}
Relative computational costs depend on the data set, particularly in
the case of the Resample-Move algorithm.
Assuming an average number of dipoles of 1, an average
number of neighbours of 25 and an average lifetime of 30 time points,
the Resample-Move algorithm has a computational cost that is
approximately 16 times
higher than the bootstrap, while the conditional importance sampling is
approximately 25 times
more costly than the bootstrap, when run with the same number of particles.

As is usual in filtering settings, the various static parameters (those which
do not change from one time point to another) are assumed known and
fixed. These parameters include the noise variance, $\Sigma_{\mathrm
{noise}}$, and
the probability of dipole birth and death, $P_{\mathrm{birth}}$ and
$P_{\mathrm{death}}$. Approaches to specifying the physical parameters are
described in the previous section and in the experimental sections
which follow.
\section{Simulation experiments}
\label{secsimulation}

In this section simulated data is used to validate and assess the performance
of the proposed method.

In simulation 1 a set of synthetic data is explicitly designed to provide
meaningful quantitative measure of performances; we used this set of
data to compare
the performances of the Resample-Move algorithm with those of the
standard bootstrap filter for
the Static model; we also provide an additional comparison with the
algorithms implementing the Random-Walk model.

In simulation 2 we apply the particle filters to a more realistic data
set and provide a visual comparison with
the estimates obtained by well-known, state-of-the-art methods.

\subsection{Simulation 1}

We first describe the generation of the synthetic data. Then we
propose a set of estimators for extracting relevant information
based on the approximation to the
posterior density provided by the particle filter.
Finally, we present a number of measures for evaluating
discrepancies between the estimated and the target dipole configuration.

\subsubsection{Generation of synthetic data}

100 different data sets were produced, according to the following protocol:
\begin{longlist}[1.]
\item[1.] The synthetic magnetic field is generated from static dipolar
sources through
the assumed forward matrix (which is taken to be the same as is used
in the model);
dipoles used to produce the synthetic data set belong to the
grid mentioned in Section \ref{secfiltering} and will be referred to as \textit{target}
dipoles.
\item[2.] Each data set comprises 70 time points and contains the activity
of 5 sources
overall; sources appear one at a time, at regular intervals of 5 time points.
\item[3.] Source locations are random, with uniform distribution in the
brain, with the constraint that no two sources in the same data set
can lie within 3 centimetres of one another.
The strength of the signal
produced by a source depends on the distance of the source from the sensors,
so that
randomness of source location implies that the signal strength---and
eventually the detectability of a source---is itself random.
\item[4.] Source orientations are first drawn at random, with uniform distribution
in the unit sphere, and then projected along the plane orthogonal to the
radial direction at the dipole location; by ``radial direction'' we
mean the direction of
the segment connecting the dipole location to the center of the sphere
best approximating the
brain surface.
Radial dipoles in a spherical conductor do not produce a magnetic field
outside of the conductor [\citet{sa87}], so this projection avoids the
creation of
undetectable
sources among the target dipoles.
%
\begin{figure}

\includegraphics{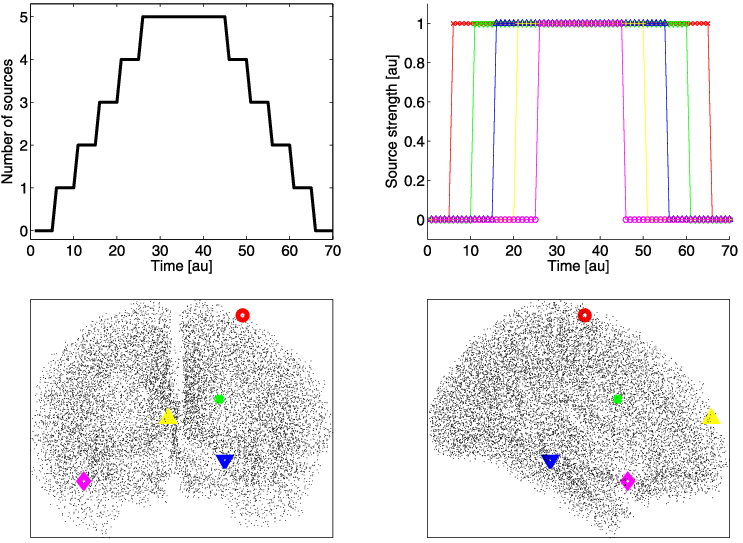}

\caption{A sample dipole configuration generating one of the 100 data
sets: in
top left panel, the number of active sources as a function of time; in
the top right panel, individual source waveforms; in the
lower panels, the source locations and the grid points, randomly drawn from
the uniform distribution in the brain.}\label{fig1}
\end{figure}
\item[5.] The intensity of the dipole moment is kept fixed throughout the lifetime
of each source, as shown in Figure \ref{fig1} (although fixed intensity clearly
does not mimic
the actual behaviour of neural sources, we adopt this simple condition
as it
helps to provide a definite measure of the number of active sources at
any time).
\item[6.] Noise is additive, zero-mean and Gaussian.
\end{longlist}

\subsubsection{Point estimates for the multi-dipole configuration}
The posterior distribution of a point process is a multidimensional
object that
is not easy to investigate and is hard to represent faithfully by
point estimates, a problem which is well known in the multi-object tracking
literature [see, e.g., \citet{vosido05} for another setting in
which a very
similar problem arises]. At the same time, often in practical
applications one is actually
interested in point
estimates; here, we are particularly interested in
evaluating whether the particle filters provide good estimates of the dipole
configuration that produced the
synthetic
data. We therefore propose a set of quantities
that can be used for this purpose, bearing in mind that they are only
low-dimensional projections of the actual output of the particle filter:
\begin{itemize}
\item The number of active sources
can be represented {via}
the marginal distribution for the number
of dipoles,
which
can be computed from the approximation to the posterior density
as
%
\begin{equation}
P(N_t=k |b_{1:t}) = \sum_{i}
w_t^i \delta_{k, N_t^i}. \label{postN}
\end{equation}
\item The location of the active sources
can be represented {via}
the intensity measure of the
corresponding
point
process.
This
provides information about the dipole location which is
highly suited to visual inspection; from the particle filter
we get the following approximation to the intensity measure:
%
\begin{equation}
\label{intensity} p(r_t|b_{1:t}) \simeq\sum
_i w_t^i \sum
_{k=1}^{N_t^i} \delta_{r_t,
r_t^{i,(k)}}.
\end{equation}
In our implementation
this
approximation is defined for all the
locations on the grid,
but not continuously over the entire volume.
\item The direction and the intensity of the estimated dipoles, that
is, the
vectorial mark of the point process: one way to provide such
information is
to calculate the average dipole moment at each location, conditional
on having a dipole at that location:
%
\begin{equation}
\label{moment} E[q_t |r] = \sum_i
w_t^i \sum_{k=1}^{N_t^i}
q_t^{i,(k)} \delta_{r, r_t^{i,(k)}}.
\end{equation}
\end{itemize}

We use the following procedure to obtain a ``representative
set'' of dipoles from the approximated posterior distribution:
\begin{longlist}[1.]
\item[1.] estimate the number of dipoles in the set by taking the mode of the
posterior distribution (\ref{postN});
\item[2.] find the $N$ highest \textit{peaks}
of the intensity measure
(\ref{intensity}): a \textit{peak} is a grid point where the
intensity measure is higher than that of its neighbours;
we take these
peaks
as point estimates of the
dipole locations;
\item[3.] for each estimated dipole location, the estimated dipole moment
will be
the average dipole moment at that location, as in (\ref{moment}).
\end{longlist}
As an alternative to the optimization in step (2), we also tried a
clustering approach
based on a Gaussian mixture model augmented with a uniform component,
devised to model possible outliers;
in the simulations below, the two approaches produced essentially the
same results (not shown).
While these measures are only low-dimensional projections and cannot
replace the rich information
contained in the posterior distribution, we feel they capture the most
important features relevant to the neuroscientist.

\subsubsection{Discrepancy measures}
Once this typical set has been estimated, the discrepancy between the estimated
dipole set $\widehat{j}_t = \{\widehat{d}_t^{(1)},\ldots,\widehat
{d}_t^{(\widehat{N}_t)} \}$ and
the target dipole set $j_t = \{d_t^{(1)}, \ldots, d_t^{(N_t))} \}$ can be
computed. However, measuring the distance between two point sets is a nontrivial
task even in the simple case when the two sets contain the same number
of points, and it becomes even more complicated
when the two sets contain a different number of points. Furthermore, in the
application under study the points also have marks, or dipole moments,
which should be taken into account.
In the following, we list several useful measures of discrepancy
between the target and the estimated dipole
configurations:

\begin{itemize}

\item Average distance from closest target (ADCT): At first we may be
interested in answering this question: how far, on average, is the estimated
dipole from any of the target dipoles? To answer this question, we can
calculate the ADCT,
defined as
%
\begin{equation}
\operatorname{ADCT}(t) = \frac{1}{\widehat{N}_t} \sum_{k=1}^{\widehat{N}_t}
\min_j \bigl|\widehat{r}_t^{(k)} -
r_t^{(j)}\bigr|,
\end{equation}
where $|\cdot|$ denotes the Euclidean norm.

\item Symmetrized distance (SD): We may also want to incorporate in the
distance measure the presence of undetected sources. To do this, we
calculate a
symmetrized version of the ADCT,
%
\begin{equation}
\operatorname{SD}(t) = \frac{1}{\widehat{N}_t} \sum_{k=1}^{\widehat{N}_t}
\min_j \bigl|\widehat{r}_t^{(k)} -
r_t^{(j)}\bigr| + \frac{1}{{N}_t} \sum
_{j=1}^{{N}_t} \min_k \bigl|\widehat
{r}_t^{(k)} - r_t^{(j)}\bigr|.
\end{equation}

\item Optimal SubPattern assignment metric (OSPA): If two estimated dipoles
are both close to the same target dipole, neither ADCT nor SD will
notice it. The OSPA metric [\citet{scvovo08}] overcomes this limitation
by forcing a one-to-one
correspondence between the estimated and the target dipoles; the OSPA metric
is defined as
%
\begin{equation}
\operatorname{OSPA}(t) = \cases{\displaystyle \min_{\pi\in\Pi_{\widehat{N}_t, N_t} } \frac{1}{\widehat{N}_t} \sum
_{k=1}^{\widehat{N}_t} \bigl|\widehat{r}_t^{(k)}
- r_t^{(\pi(k))}\bigr|, &\quad $\mbox{if } \widehat{N}_t \leq
N_t,$\vspace*{2pt}
\cr
\displaystyle\min_{\pi\in\Pi_{{N}_t, \widehat{N}_t}}
\frac{1}{{N}_t} \sum_{k=1}^{{N}_t}\bigl |
\widehat{r}_t^{(\pi(k))} - r_t^{(k)}\bigr|, &\quad $
\mbox{if } \widehat{N}_t > N_t,$}
\end{equation}
where $\Pi_{k,l}$ is the set of all permutations of $k$ elements drawn from
$l$ elements. Note that this metric only calculates the discrepancy between
the dipoles in the smaller set and the subset of dipoles in the larger set
that has the smaller discrepancy with those in the smaller set.

\item Widespread measure (WM):
Finally, we want to combine discrepancies in the source
location with discrepancies in the dipole moment. The following
measure does
this by replacing each dipole (both in the target dipole set and in the
estimated dipole set) with a Gaussian-like function in the brain, centered
at the dipole location, with fixed variance and height proportional to the
dipole moment; the difference between the two spatial distributions is then
integrated in the whole brain:
%
\begin{equation}
\operatorname{WM}(t) = \int\Biggl\llvert \Biggl[\sum_{k=1}^{\widehat{N}_t}
\bigl|\widehat{q}_t^{(k)}\bigr| \mathcal{N}\bigl(r;
\widehat{r}_t^{(k)} \sigma\bigr) - \sum
_{k=1}^{{N}_t} \bigl|{q}_t^{(k)}\bigr|
\mathcal{N}\bigl(r; {r}_t^{(k)}, \sigma\bigr) \Biggr] \Biggr
\rrvert \,dr,
\end{equation}
where the integral must in practice be approximated by numerical methods.
\end{itemize}

\subsubsection{Results}

We ran the Resample-Move particle filter on the 100 synthetic data sets
described at the beginning of this section.
We also ran a bootstrap filter on the same data to evaluate
its ability to sample the Static model's posterior.
In addition, we ran both a standard bootstrap and an improved filter, as
described in the previous section, implementing the Random-Walk model.

All filters were run with 10,000 particles. In addition, in order to
compare the
performances for approximately equal computational cost, we ran both
the Resample-Move filter and the improved filter for the Random-Walk
model with 500 particles.
All filters were run with the same parameter values: the standard
deviation of the Gaussian prior for the dipole moment was set to $\sigma
_q = 1$ nAm;
the noise covariance matrix was diagonal, with the same value $\sigma
_{\mathrm{noise}}^2$ for each channel and estimated from the first 5 time
points. This was done in analogy with typical MEG experiments with
external stimuli,
where a pre-stimulus interval is typically used to estimate the noise variance.

\begin{figure}

\includegraphics{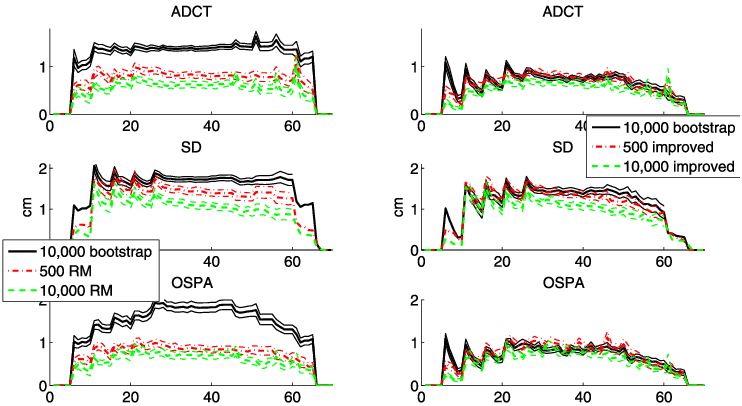}

\caption{Simulation 1. Discrepancy measures for the Static
model (left column) and the Random-Walk model (right
column).}
\label{figdiscrepancy}
\end{figure}

\begin{figure}[b]

\includegraphics{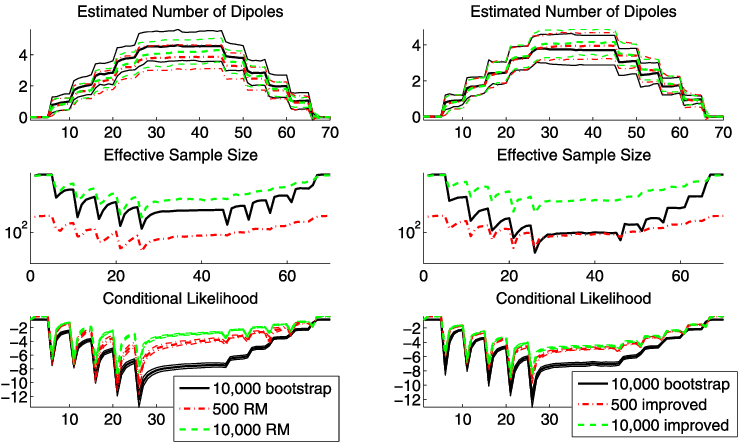}

\caption{Simulation 1. Estimated number of sources, effective sample
size and conditional
likelihood for the Static model (left column) and the
Random-Walk model (right column).}
\label{figdiagnostics}
\end{figure}

We computed the discrepancy measures proposed in Section \ref{secsimulation}. The results
are shown in Figure \ref{figdiscrepancy}; the widespread measure
provided results that are
very similar to those of the OSPA metric, hence, for brevity it is not
shown here.
In Figure \ref{figdiagnostics} we show the estimated number of
sources, the effective sample size, as given by equation (\ref{ess}),
and the conditional likelihood as a function of time.

All the discrepancy measures indicate that the Resample-Move
particle filter provides a substantial improvement over the bootstrap filter.
The use of three different measures, in conjunction with the
observation of the estimated number of sources in Figure \ref{figdiagnostics},
gives more insights about the qualitative nature of the improvements.
First of all, the ADCT indicates that the dipoles
estimated with the Resample-Move are on average much closer to the
target sources;
in addition, there is a rather small difference between the results
obtained running the Resample-Move with 10,000 particles and with 500
particles. The average localization error is about 7 mm with the new
model in contrast to the bootstrap particle filter which achieves an
average localization error of 1.4~cm.
The SD provides a slightly different result: there is more difference
here between 500 and 10,000
particles; this is due to the fact that using a higher number of particles
allows the algorithm to explore the state space better. In addition,
the relatively small difference with the bootstrap filter
here is due to the fact that the Resample-Move algorithm tends to
slightly underestimate
the number of sources, which is penalized by the SD. Finally, in terms
of the OSPA
measure, the Resample-Move provides a rather large improvement over the
bootstrap: that the difference is so large is due to the fact that the
bootstrap filter tends to
overestimate the number of dipoles, in the proximity of a target dipole
(as it is unable to update dipole locations and the likelihood strongly
supports the
presence of some additional signal source).
This does not have a big impact on the ADCT, but is highly penalized by
the OSPA.
Observation of the ESS and of the conditional likelihood in Figure
\ref{figdiagnostics} strengthens the previous results. The
Resample-Move algorithm maintains a higher ESS throughout the whole temporal
window, in which the number of dipoles increases from zero to five and then
returns to zero.
The conditional likelihood further adds to the general evidence that the
Resample-Move algorithm has better explored the state space whilst the
bootstrap algorithm has missed
a substantial part of the mass of the posterior. This plot also demonstrates
that, as one would expect, better performance is obtained with a larger number
of particles. However, with just 500 particles the Resample-Move algorithm
provides better localisation performance than the bootstrap filter with 10,000
particles---as demonstrated by the various discrepancy measures.

Finally, we compare the performance of the Static model and the Random-Walk
model. Noting that the bootstrap algorithm is unable to provide adequate
inference for this model, as shown in Figure \ref{figdiscrepancy}, we consider the proposed
Resample-Move algorithm which is designed specifically to address the
limitations of the simpler algorithm in this setting. The discrepancy measures
indicate that the two models perform rather similarly in terms of average
localization accuracy; this has to be regarded as a positive fact,
since the
localization accuracy of the Random-Walk model was already
recognized as being satisfactory [\citet{soetal09}], and the Static
model is in
fact a model for which inference is harder. However, in most experimental
conditions, the Random-Walk model is not believed to be
as physiologically plausible as the Static model.
Notably, in this synthetic experiment in which we
know that the dipoles are actually static, we observe that
the Static model leads to higher conditional likelihood than the random
walk model. As in the
context of Bayesian model selection, this implies that the data
supports the
Static model in preference to the Random-Walk model. However, some
caution should be exercised in
interpreting these results, as we are not dealing with the full Bayesian
marginal likelihood: in both cases the true static parameters (noise
variance, scale of random walk) have been fixed and so the time
integral of
these conditional likelihoods can only be interpreted as a marginal
\textit{profile} likelihood (it is not currently feasible to consider
the penalisation of more complex models afforded by
a fully Bayesian method in which the unknown parameters were
marginalized out).

\subsection{Simulation 2}

We consider a simulated data set designed to mimic a real evoked
response experiment,
with the typical bell-shaped source time courses and real noise superimposed.

\subsubsection{Generation of synthetic data}

\begin{figure}

\includegraphics{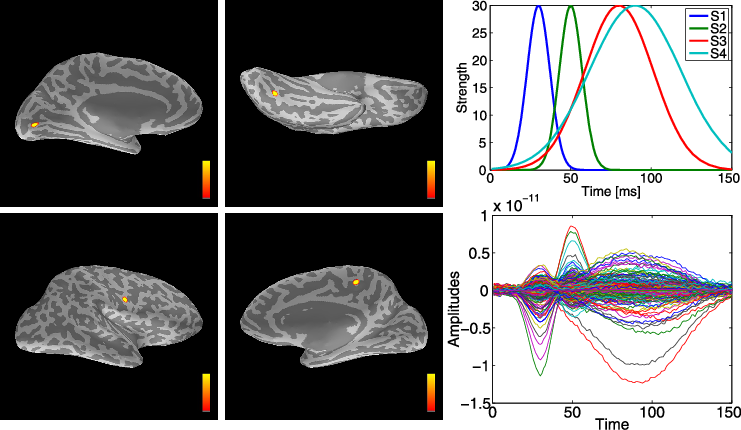}

\caption{Simulation 2. Source locations (top row \textup{S1} and \textup{S2}, bottom row
\textup{S3} and \textup{S4}), source time courses and generated noisy field.}
\label{figsim2}
\end{figure}

We generated the synthetic data shown in Figure \ref{figsim2}.
A first source (S1) in the central occipital area has peak strength at
$30$~ms after the hypothetical stimulus; a second occipital
but more lateral and ventral source (S2) peaks at $50$~ms; then one
temporo-parietal source on the lateral surface (S3)
and one parietal source on the medial surface (S4) peak at $80$ and
$90$~ms, respectively, with a substantial temporal overlap.
Noise free measurements were generated from the sources displayed in
Figure \ref{figsim2} through the lead field
matrix. In order to mimic a real-world data set, these noise-free data
were added to the pre-stimulus signal from a real experiment,
involving the same subject that was used to create the lead field
matrix. The resulting noisy data are shown in the same figure.

\subsubsection{Comparison with other methods}

We compare the Resample-Move particle filter implementing the Static
model with the particle filter implementing the Random-Walk model,
as well as with three state-of-the-art methods for MEG source estimation:
\begin{itemize}
\item the recursively applied multiple signal classification
(RAP-MUSIC) algorithm [\citet{molele92,mole99}] is perhaps the most popular
method for automatic estimation of current dipole parameters from MEG
data, and is widely used as a reference method for both MEG and EEG
dipole modeling [\citet{hoetal12,wuetal12}]. After selection of a
suitable number of components to identify a \textit{signal subspace}, RAP-MUSIC computes
an index at each point within the brain, representing the subspace
correlation [\citet{gova84}] between the leadfield at that point and the
signal subspace. Peaks of this function are often used as point
estimates of dipole locations;
\item dynamic Statistical Parametric Mapping (dSPM) [\citet{daetal00}]
is a well-known method, whereby Tikhonov regularization is applied at
each time point independently, and the so-obtained estimate of the
electrical current distribution is divided by a location-dependent
estimate of the noise variance; the resulting quantity, also named
\textit{activity estimate}, has a $t$-distribution under the null
hypothesis of no activity. In the experiments below, we used the dSPM
algorithm contained in the MNE software;
\item$L_1L_2$ [\citet{ouhago09}, Gramfort, Kowalski and\break H{\"{a}}m{\"{a}}l{\"{a}}inen (\citeyear{grkoha12})] is a spatio-temporal
regularization method, whereby the penalty term has a mixed norm: an L1
norm in the
spatial domain, encouraging sparsity of the estimated current
distribution, and an L2 norm in the temporal domain, encouraging
continuity of the source waveforms. In the experiments below, we used
the $L_1L_2$ algorithm contained in the EMBAL Matlab toolbox
(\url{http://embal.gforge.inria.fr/}).
\end{itemize}
We notice that the proposed comparison is necessarily a comparison of
nonhomogeneous methods: while RAP-MUSIC is fundamentally a dipole
localization method, $L_1L_2$ produces estimates of a continuous
current distribution, and dSPM provides a statistical measure
of activity; the particle filters, on the other hand, produce a dynamic
approximation to the posterior distribution for a multiple
current dipole model. For these reasons, a quantitative comparison
resembling that of the previous section would be questionable
and would fail to illustrate the fundamental differences between these methods.
Therefore, in the following we provide a visual comparison of the main
ouputs: the posterior intensity function for
the particle filters; the subspace correlation index for RAP-MUSIC; the
activity estimate for dSPM; the electrical current estimate
for $L_1L_2$.

\begin{figure}

\includegraphics{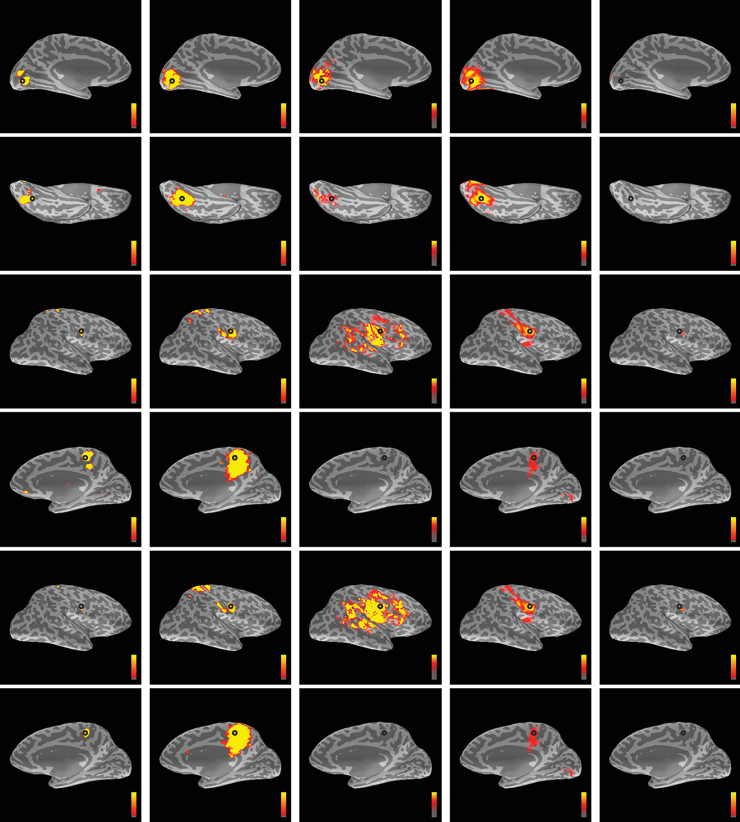}

\caption{Simulation 2. Comparison of the maps produced by the Static
model particle filter (first column), the Random-Walk model particle
filter (second column), dSPM (third column), RAP-MUSIC (fourth column)
and $L_1L_2$ (fifth column) at different time points: 30~ms (first row),
50~ms (second row), 75~ms (third and fourth rows), 90~ms (fifth and
sixth rows). Black circles mark the locations of the true sources.}
\label{figsim2res}
\end{figure}

\subsubsection{Results}

In Figure \ref{figsim2res} we show the results provided by the
different methods at selected time points. Light and dark grey represent
here the anatomical details, while color represents the estimated
quantities, with values increasing from a threshold, red, to a maximum
value, yellow. The color scale is different for each method: for the
particle filters the threshold is $10^{-5}$ and the maximum is $0.1$;
for dSPM the threshold is $5$ and the maximum is $15$; for RAP-MUSIC
the threshold is $70$ and the maximum is $100$; for $L_1L_2$ the
threshold is $1$ and the maximum is~$30$. We notice that the
correlation index provided by RAP-MUSIC is in fact not time-dependent;
it is only for presentation purposes that we repeat the same
figures at rows 3--4 and rows 5--6. The visualization of the brain is
also worth a brief comment. The smooth brain surface in these images
is indeed a computer representation in which the highly folded cortical
surface is ``inflated'' in such a way that the activity in the
sulci becomes easily visible. As a consequence, spatially adjacent
volumes---for example, the portion of the cortex in two adjacent sulci---may
be moved apart, therefore, the presence of multiple close-by peaks or
blobs in these images is most often an artefact due to the
visualization, rather than an actual multi-modality of the
three-dimensional spatial distribution of the displayed quantity.

At $t=30$~ms, source 1 is recovered by all methods, the $L_1L_2$
solution being only slightly more superficial than the actual source;
this happens despite the use of the same depth weighting method
proposed in \citet{lietal06} and described earlier in this paper. A
direct comparison of the Static model with the Random-Walk model
illustrates that the Static model tends to produce sparser probability
maps: this is due to the Resample-Move algorithm being able to
accumulate information on the source location with time.
At $t=50$~ms, source 2 is correctly recovered by all methods, except
$L_1L_2$; $L_1L_2$ does not produce any detectable output in the ventral
area where source 2 is. In fact, we were able to estimate this source
with $L_1L_2$ by modifying the value of the regularization parameter,
but this came at the cost of making the solution at all time points
notably less sparse, and much more similar to the one provided
by dSPM.
At $t=75$~ms, source 3 is recovered by all methods, with $L_1L_2$
providing a particularly accurate and sparse solution. However,
source 4 is not recovered by $L_1L_2$ nor by dSPM; tweaking the
parameters did not work for either method. On the other hand, the subspace
correlation computed by RAP-MUSIC does have a local maximum around
source 4, but its value of about 0.8 is not ``close to 1,'' hence, whether
source 4 will be detected depends on the subjective choice of a threshold.
The same comment applies to $t=90$~ms, where, in addition, we notice
that, as already noted, the Static model implemented in the
Resample-Move particle filter produces a more focal posterior map as
time goes on, as a consequence of the accumulation of information
on the source, while the Random-Walk model does not.

\section{Application to real data}
\label{secexample}

We applied the Resample-Move algorithm to real MEG recordings
which were
obtained during stimulation of a large nerve in the arm. This choice is
motivated by the
fact that the neural response to this type of somatosensory stimulation is
relatively simple and rather well understood [\citet{maetal97}], and therefore
allows a meaningful assessment of performance.

\subsection{Experimental data}

We used data from a Somatosensory Evoked Fields (SEFs) mapping experiment.
The recordings were performed after informed consent was\vadjust{\goodbreak} obtained, and had
prior approval by the local ethics committee.
Data were acquired with a
306-channel MEG device (Elekta Neuromag Oy, Helsinki, Finland)
comprising 204
planar gradiometers and 102 magnetometers in a helmet-shaped
array. The left median nerve at the wrist was electrically
stimulated at the motor threshold with an interstimulus
interval randomly varying between 7.0 and 9.0 s. The
MEG signals were filtered to 0.1--200 Hz and sampled at
1000 Hz. Electrooculogram (EOG) was used to monitor eye movements
that might produce artefacts in the MEG recordings; trials with
EOG or MEG exceeding 150 mV or 3 pT/cm, respectively, were excluded and
84 clean trials
were averaged. To reduce external interference, the signal space
separation method [\citet{taetal04}] was applied
to the average.
A 3D digitizer and four head position indicator coils
were employed to determine the position of the subject's
head within the MEG helmet with respect to anatomical
MRIs obtained with a 3-Tesla MRI device (General Electric,
Milwaukee, USA).

\subsection{Results}

The Resample-Move particle filter implementing the Static model
was applied to the MEG recordings; for the sake of comparison, we also
applied the
particle filter based on the Random-Walk model and the conditional
sampling, as well as
dSPM, RAP-MUSIC and $L_1L_2$, the methods already used and briefly
described in simulation 2.
Both particle filters were run with the same parameter values.
The standard deviation of the Gaussian prior for the dipole moment was
set to $\sigma_q =
50$ nAm, which is a physiologically plausible value; varying the value
of this parameter
did not qualitatively alter the reconstructions obtained.
The noise covariance matrix was diagonal, the diagonal entries assuming
either of two values,
one for gradiometers and one for magnetometers; these values were
obtained by averaging, across
homogeneous channels, the channel-specific estimates of the standard
deviation obtained from the pre-stimulus
interval.

\begin{figure}

\includegraphics{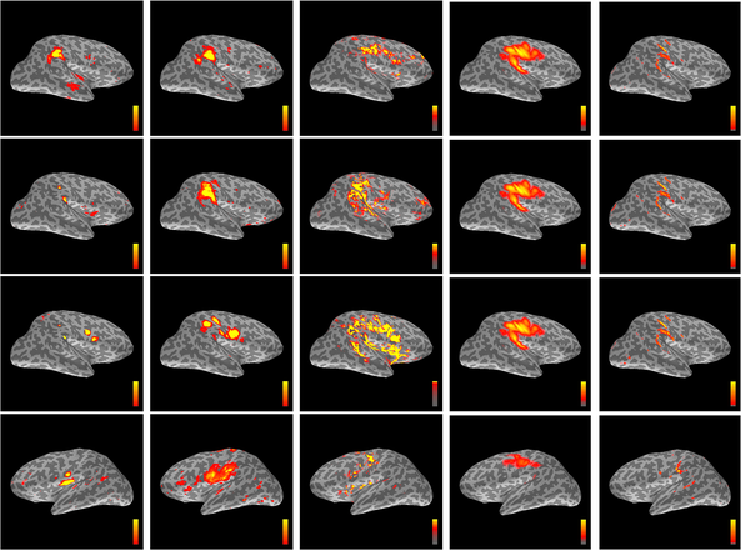}

\caption{SEF data. Comparison of the maps produced by the Static model
particle filter (first column), the Random-Walk model particle filter
(second column), dSPM (third column), RAP-MUSIC (fourth column) and
$L_1L_2$ (fifth column) at different time points: 25~ms (first row),
50~ms (second row), 85~ms (right hemisphere on the third row, left
hemisphere on the fourth row).}
\label{figrealdata}
\end{figure}

Figure \ref{figrealdata} illustrates the localization provided by the five
methods. We show snapshots at three time points.
The very first response in the primary somatosensory area SI, at 25~ms, is
localized by both the Static and the Random-Walk particle filters in a very
similar way. The correlation index in RAP-MUSIC---which we recall is
not time varying---clearly has a peak
around the same location; the $L_1L_2$ activity estimate is slightly
more superficial but still very close, while
the dSPM estimate is rather widespread and indicates activity in
slightly more frontal areas.
At time $t=50$~ms after stimulus, the Static and the Random-Walk model
are showing the same behaviour
already observed in simulation 2: as the SI area has been active for
the past 25~ms, the posterior map of the Static model is much more
concentrated now, having accumulated information on the source
location; the
Random-Walk model indicates activity in the same area but provides a more
blurred image. The estimate of dSPM is now closer to the probability
maps provided by the two filters,
while $L_1L_2$ does not show significant changes from the previous snapshot.
At time $t=85$~ms, finally, we observe more activation in SI, and the additional
activity in the ipsilateral and contralateral SII: observing the
posterior maps
provided by the Static model we observe, as in \citet{maetal97}, that the
contralateral SII activation is more frontal than the ipsilateral SII
activation. The Random-Walk model provides, again, more blurred images.
The dSPM estimate is again more widespread.
RAP-MUSIC has local maxima around 0.85 in a similar area as the
particle filters for the right hemisphere, while there is a
slight disagreement on the left hemisphere; finally, the source
distribution in $L_1L_2$ is not much changed on the right hemisphere,
while on the left hemisphere it provides a slightly more posterior
localization with respect to the one provided by the particle filter.

\begin{figure}

\includegraphics{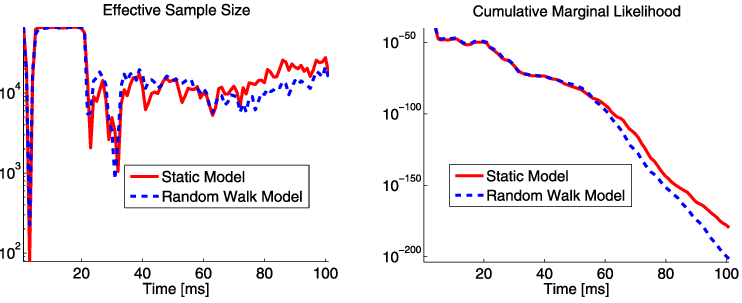}

\caption{The effective sample size and the marginal likelihood, as obtained
with the Static and the Random-Walk model with the SEF data.}
\label{figrealdatadiag}
\end{figure}

\begin{figure}

\includegraphics{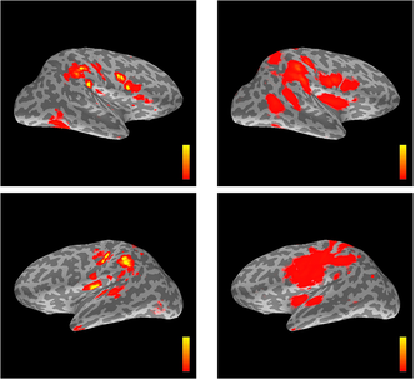}

\caption{Time-integrated probability maps: the Static model (left column)
exhibits less temporal variation than the Random-Walk model (right
column).}
\label{figrealdataglobal}
\end{figure}

In Figure \ref{figrealdatadiag} we show the cumulative marginal
likelihood (\ref{eqmarginallikelihood}) and the
effective sample size for the two particle filters. While the effective
sample size produces rather similar results for the two models,
the marginal likelihood indicates that after approximately $t=60$~ms
the Static model provides higher likelihood than the Random-Walk
model.
Importantly, the cumulative likelihood at time $t$ is the likelihood of
the whole time series \textit{up to} time $t$.
The fact that the difference between the two models tends to increase
with time indicates that, as more data are gathered,
the Static model is increasingly preferentially supported by the
measurements. The ratio of the two likelihoods at the terminal time point
indicates that the whole time series is several orders of magnitude
more likely under the Static model than under the Random-Walk model,
thus providing confirmation that the Static model is a much better
representation of the underlying neurophysiological
processes than the Random-Walk model.
An additional point that deserves to be highlighted here is that not only
are the probability maps provided by the Static model sparser than those
provided by the Random-Walk, but also (as one might reasonably expect)
they show less temporal variation.
To illustrate this point, in Figure \ref{figrealdataglobal} we
show two maps that have been obtained by integrating over time the dynamic
probability maps provided by the Static and the Random-Walk filters:
while the Static model
has high probability in few small areas and negligible probability elsewhere,
the Random-Walk model provides a flatter image, with rather homogeneous
probability values in a larger area,
a consequence of the fact that the Random-Walk model attaches a large
part of its posterior probability mass to dipoles which move around the brain.

As we run several independent realizations of the filters with the same
parameter values and 100,000 particles, we observed that for $t> 75$~ms
not all
the runs provide exactly the same estimates.
While in the majority of the runs the mode of the posterior distribution
consistently presents the source configuration depicted in Figure
\ref{figrealdata}, in approximately 10\% of the runs the ipsi-lateral
and contra-lateral
SII sources are replaced with a pair of sources in between the two
hemispheres, one at the
top in the motor area and one rather deep and central; the two SII
areas are still represented in the posterior distribution, but with
slightly lower mass,
and may not appear in low-dimensional summaries of the posterior. As noted
previously, accurately summarising high-dimensional multi-object
posteriors is
known to be a rather difficult problem.
Finally, we note that if too small a sample size was employed,
then we found that the quality of the approximation of the posterior
could deteriorate to the
point that the posterior did not contain mass in a neighbourhood of the
configuration shown in Figure \ref{figrealdata}. Naturally, sample-based
approximations to high-dimensional distributions fail to accurately capture
the important features of those distributions if the sample is too small.
In the case of a sample of size 10,000 we observed this failure mode in
less than 10\% of the runs. In
practice, such a phenomenon should be easily detected by replicating the
estimation procedure a number of times.

\section{Discussion}
\label{secDiscussion}

In this study we have presented a new state-space model for dynamic
inference on current dipole parameters from MEG data with particle filtering.
The model has been devised to reflect the common neurophysiological
interpretation of a current dipole as the activity of a small patch of cortex:
the number of dipoles is time-varying, as dipoles can appear and
disappear, and
dipole locations are fixed during the dipole life time.
Standard sequential Monte Carlo algorithms are not well suited to
``filtering'' of static parameters; for the same reasons simple sequential
importance resampling is not able to efficiently approximate the posterior
distributions associated with these\vadjust{\goodbreak} near-static objects. We have
developed a
Resample-Move type algorithm with carefully designed proposal distributions
which are appropriate for inference in this type of model.

We have used synthetic data to show that the average localization error provided
by the Resample-Move algorithm is close to 5 mm, that is, the average
grid spacing,
even when the data are produced by five simultaneous dipoles. In
addition, the effective sample size
remains high even when the filter explores the high-dimensional
parameter space
of five dipoles, consistent with a good approximation of the posterior
distribution.
Although the quality of the approximation naturally depends on the
sample size,
we demonstrated good results can be obtained with a realistic
computational cost.
Finally, comparison of the conditional likelihood of our Static dipole
model with that of a Random-Walk
model indicates that the proposed method is actually capable of
providing a
better explanation of the data.

We have used a second simulation study to assess the localization
capability of the particle
filter in comparison with dSPM, RAP-MUSIC and $L_1L_2$. The activity
maps showed that both particle filters
were able to identify all the four sources in the simulation, while
dSPM and $L_1L_2$ missed at least one,
and RAP-MUSIC provided local peaks but with low intensity for one
source. In addition, comparison of the
probability maps provided by the Static and the Random-Walk models
shows that the Static model coherently
accumulates information on the source and provides more focal maps with
time, while the Random-Walk does not.

Application of the proposed method to an experimental data set has produced
similar results: the effective sample size and the conditional
likelihood remain
high throughout the whole time series; the posterior probability maps
are well
in accordance with what is understood to be the usual brain response to median
nerve stimulation.
The Static model leads to physiologically-interpretable output which is
consistent with the biomedical understanding of the dipole model. We
did not
observe the type of artefacts found with the Random-Walk model in which
dipoles slowly moved across the brain surface when using this model.

Future research will concentrate on increasing the number of samples and
decreasing the computational time. Implementation on GPUs should provide
a viable way to reduce the computational time exploiting massive
parallelization and given performance improvements observed in similar
settings [\citet{algorithmsmontecarloLYGDH10}], thereby facilitating real-time
implementation. This together with the bounded per-iteration computational
cost of the filtering algorithm is a significant motivation of the approach.
Improving the efficiency of the MCMC step is also of interest.
Other possible interesting research directions include the use of smoothing
[\citet{brdoma10}] techniques and estimation of the static parameters
(which were here fixed a priori
using approaches prevalent in the literature) both online [\citet
{kaetal09}] and offline [\citet{mcmcmethodologyADH10},
\citet{smcmethodologyCJP11}], as well as generalization of the source model.

\section*{Acknowledgments}
We gratefully acknowledge the help of Dr. Lauri Park\-konen and Dr.
Annalisa Pascarella, who together with A.~Sorrentino undertook
collection, post-processing and analysis of the experimental data, and
of Dr. Alexandre Gramfort,
who kindly provided support on the use of the EMBAL Matlab package.

%


\printaddresses

\end{document}